\documentclass[onecollarge]{svjour2}       
\usepackage{graphicx}
\def\geqsim{\mathbin{\;\raise1pt\hbox{$>$}\kern-8pt\lower3pt\hbox{$\sim$}\;}}
\def\leqsim{\mathbin{\;\raise1pt\hbox{$<$}\kern-8pt\lower3pt\hbox{$\sim$}\;}}

\begin{document}

\title{Is the evidence for dark energy secure?}
\author{Subir Sarkar}
\institute{Rudolf Peierls Centre for Theoretical Physics, University
of Oxford, 1 Keble Road, Oxford OX1 3NP, UK \\
\email{s.sarkar@physics.ox.ac.uk} }

\date{Received: 22 October 2007 / Accepted: 23 October 2007}

\maketitle

\begin{abstract}
  Several kinds of astronomical observations, interpreted in the
  framework of the standard Friedmann--Robertson--Walker cosmology,
  have indicated that our universe is dominated by a Cosmological
  Constant. The dimming of distant Type Ia supernovae suggests that
  the expansion rate is accelerating, as if driven by vacuum energy,
  and this has been indirectly substantiated through studies of
  angular anisotropies in the cosmic microwave background (CMB) and of
  spatial correlations in the large-scale structure (LSS) of
  galaxies. However there is no compelling {\em direct} evidence yet
  for (the dynamical effects of) dark energy.  The precision CMB data
  can be equally well fitted without dark energy if the spectrum of
  primordial density fluctuations is not quite scale-free and if the
  Hubble constant is lower globally than its locally measured
  value. The LSS data can also be satisfactorily fitted if there is a
  small component of hot dark matter, as would be provided by
  neutrinos of mass $\sim0.5$ eV. Although such an Einstein--de Sitter
  model cannot explain the SNe~Ia Hubble diagram or the position of
  the `baryon acoustic oscillation' peak in the autocorrelation
  function of galaxies, it may be possible to do so e.g. in an
  inhomogeneous Lemaitre--Tolman--Bondi cosmology where we are located
  in a void which is expanding faster than the average. Such
  alternatives may seem contrived but this must be weighed against our
  lack of any fundamental understanding of the inferred tiny energy
  scale of the dark energy. It may well be an artifact of an
  oversimplified cosmological model, rather than having physical
  reality.

\end{abstract}

\section{Introduction}

Following his formulation of general relativity Einstein
\cite{Einstein:1917} boldly applied the theory to the universe as a
whole. The first cosmological model was {\em static} to match the
known universe, which at that time was restricted to the Milky way,
and to achieve this Einstein introduced the `cosmological constant'
term (for a historical perspective, see
\cite{Straumann:2002he}). Within a decade however Slipher and Hubble
demonstrated that the nebulae on the sky are in fact other `island
universes' like the Milky Way and that they are mainly receeding from
us --- the universe is expanding. Einstein wrote to Weyl in 1933: {\em
  ``If there is no quasi-static world, then away with the cosmological
  term''}.

This however is not a matter of choice since general coordinate
invariance, which Einstein's equation is based on, permits an
arbitrary constant (multiplied by the metric tensor) to be added to
the lhs:
\begin{equation}
R_{\mu\nu} - \frac{1}{2}g_{\mu\nu}R + \lambda_\mathrm{metric} g_{\mu\nu} 
 = \frac{-T_{\mu\nu}}{M_\mathrm{P}^2}.
\label{gr}
\end{equation}
Here we have written Newton's constant, $G_\mathrm{N} \equiv
1/8\pi\,M_\mathrm{P}^2$ where $M_\mathrm{P} \simeq
2.4\times10^{18}$~GeV is the (reduced) Planck mass in natural units
($\hbar = k_\mathrm{B} = c = 1$). With the subsequent development of
quantum field theory it became clear that the energy-momentum tensor
on the rhs can also be freely scaled by another additive constant
multiplying the metric tensor, which reflects the (Lorentz invariant)
energy density of the vacuum:
\begin{equation}
\langle T_{\mu\nu}\rangle_{\rm fields} 
 = -\langle\rho\rangle_{\rm fields}g_{\mu\nu}.
\end{equation}
This contribution from the matter sector adds to the ``bare'' term
from the background geometry, yielding an effective cosmological
constant:
\begin{equation}
\Lambda = \lambda_\mathrm{metric} 
          + \frac{\langle\rho\rangle_{\rm fields}}{M_\mathrm{P}^2},
\label{tuning}
\end{equation}
or, correspondingly, an effective vacuum energy:
\begin{equation}
\rho_\mathrm{v} \equiv \Lambda M_\mathrm{P}^2. 
\label{rhov}
\end{equation}

Einstein {\em assumed} without any observational evidence that the
universe is perfectly homogeneous. We know that the universe is quite
isotropic about us so this is in fact likely if we are not in a
special location --- an assumption later dignified by Milne as the
`Cosmological Principle'. Then using the {\em maximally symmetric}
Robertson-Walker metric to describe space-time
\begin{equation}
\mathrm{d}s^2 \equiv g_{\mu\nu} \mathrm{d}x^\mu \mathrm{d}x^\nu 
 = \mathrm{d}t^2 - a^2(t)[ \mathrm{d}r^2/(1 - kr^2) + r^2 \mathrm{d}\Omega^2],
\label{rw}
\end{equation}
we obtain the Friedmann equations describing the evolution of the
cosmological scale-factor $a (t)$:
\begin{equation}
H^2 \equiv \left(\frac{\dot{a}}{a}\right)^2 =
\frac{\rho}{3M_\mathrm{P}^2} - \frac{\kappa}{a^2} + \frac{\Lambda}{3}, \qquad
\frac{\ddot{a}}{a} = -\frac{1}{6M_\mathrm{P}^2}(\rho + 3p) + \frac{\Lambda}{3},
\label{fried}
\end{equation}
where $\kappa = 0, \pm1$ is the 3-space curvature signature and for
ordinary matter (`dust') and radiation we have used the `ideal gas'
form: $T_{\mu\nu} = p g_{\mu\nu} + (p + \rho) u_\mu u_\nu$, with
$u_\mu \equiv \mathrm{d}x_\mu/\mathrm{d}s$. The conservation equation
$T^{\mu\nu}_{;\nu} = 0$ implies $\mathrm{d}(\rho a^3)/\mathrm{d}a = -3
p a^2$, so given the `equation of state parameter' $w \equiv p/\rho$,
the evolution history can now be constructed. Since the redshift is $z
\equiv a/a_0 - 1$, for non-relativistic particles with $w \simeq 0$,
$\rho_\mathrm{NR} \propto (1 + z)^{-3}$, while for relativistic
particles with $w = 1/3$, $\rho_\mathrm{R} \propto (1 + z)^{-4}$, but
for the cosmological constant, $w = -1$ and $\rho_\mathrm{v} =
$constant. Thus radiation was dynamically important only in the early
universe (for $z \geqsim 10^4$) and for most of the expansion history
only non-relativistic matter is relevant. The Hubble equation can be
rewritten with reference to the present epoch (subscript 0) as
\begin{eqnarray}
H^2 &=& H_0^2 \left[\Omega_\mathrm{m} (1 + z)^3 + \Omega_\kappa (1 + z)^2 
+ \Omega_\Lambda \right], \\ 
\Omega_\mathrm{m} &\equiv& \frac{\rho_{\mathrm{m}0}}{\rho_\mathrm{c}}, \quad
\Omega_\kappa \equiv -\frac{\kappa}{a_0^2 H_0^2}, \quad
\Omega_\Lambda \equiv \frac{\Lambda}{3 H_0^2},
\label{hub}
\end{eqnarray}
where $\rho_\mathrm{c} \equiv 3H_0^2M_\mathrm{P}^2/8\pi \simeq (3 \times
10^{-12} \,\mathrm{GeV})^4 h^2$ is the `critical density' and the
present Hubble parameter is $H_0 \equiv
100h~\mathrm{km}~\mathrm{s}^{-1}~\mathrm{Mpc}^{-1}$ with $h \simeq
0.7$, i.e. about $10^{-42}$~GeV.  This yields the sum rule
\begin{equation}
\Omega_\mathrm{m} + \Omega_\kappa + \Omega_\Lambda = 1,
\label{sumrule}
\end{equation}
so cosmological models can be usefully displayed on a `cosmic
triangle' \cite{Bahcall:1999xn}.

As emphasised in an influential review \cite{Weinberg:1988cp}, given
that the density parameters $\Omega_\mathrm{m}$ and $\Omega_\kappa$
were observationally constrained already to be not much larger than
unity, the two terms in eq.(\ref{tuning}) are required to somehow {\em
  conspire} to cancel each other in order to satisfy the approximate
constraint
\begin{equation}
|\Lambda| \leqsim H_0^2,
\end{equation}
thus bounding the present vacuum energy density by $\rho_\mathrm{v}
\leqsim 10^{-47}\,\mathrm{GeV}^4$ which is a factor of over $10^{120}$
below its ``natural'' value of $\sim M_\mathrm{P}^4$ --- the
`cosmological constant problem'. Subsequently, several types of
evidence have been advanced to argue that this inequality is in fact
saturated with $\Omega_\Lambda \simeq 0.7$ ($\Rightarrow \Lambda
\simeq 2H_0^2$), $\Omega_\mathrm{m} \simeq 0.3$, $\Omega_\kappa \simeq
0$ (see \cite{Sahni:1999gb,Peebles:2002gy}), i.e. there is non-zero
vacuum energy of {\em just the right order of magnitude so as to be
  detectable today}.

In the Lagrangian of the Standard $SU(3)_\mathrm{c} \otimes
SU(2)_\mathrm{L} \otimes U(1)_Y$ Model (SM) of electroweak and strong
interactions, the term corresponding to the cosmological constant is
one of the two `super-renormalisable' terms allowed by the gauge
symmetries, the second one being the quadratic divergence in the mass
of fundamental scalar fields due to radiative corrections (see
\cite{Zwirner:1995iw}). To tame the latter sufficiently in order to explain
the experimental success of the SM has required the introduction of a
`supersymmetry' between bosonic and fermionic fields which is `softly
broken' at about the Fermi scale,
$M_\mathrm{EW}\sim\,G_\mathrm{F}^{-1/2}\simeq246$~GeV Thus the cutoff
scale of the SM, viewed as an effective field theory, can be lowered
from $M_\mathrm{P}$ down to $M_\mathrm{EW}$, at the expense of
introducing over 100 new parameters in the Lagrangian, as well as
requiring delicate control of the non-renormalisable operators which
can generate flavour-changing neutral currents, nucleon decay etc, so
as not to violate experimental bounds. This implies a {\em minimum}
contribution to the vacuum energy density from quantum fluctuations of
${\cal O}(M_\mathrm{EW}^4)$, i.e. halfway on a logarithmic scale
down from $M_\mathrm{P}$ to the energy scale of ${\cal
O}(M_\mathrm{EW}^2/M_\mathrm{P})$ corresponding to the observationally
indicated vacuum energy. Thus even the introduction of supersymmetry
cannot eradicate a discrepancy by a factor of at least $10^{60}$
between the natural expectation and observation.

It is likely that a satisfactory resolution of the cosmological
constant problem can be achieved only in a quantum theory of
gravity. Recent developments in string theory and the possibility that
there exist new dimensions in Nature have generated many interesting
ideas concerning possible values of $\Lambda$ (see
e.g. \cite{Witten:2000zk,Padmanabhan:2002ji,Nobbenhuis:2004wn}). Nevertheless
it is the case that there is no accepted solution to the enormous
discrepancy discussed above. The problem is not new but there has
always been the hope that some day we would understand why $\Lambda$
is exactly zero, perhaps due to a new symmetry principle. However if
it is in fact non-zero and dynamically important {\em today}, the
crisis is even more severe since this also raises a cosmic
`coincidence' problem, viz. why is the present epoch
special?~\footnote{While $\Lambda \sim H_0^2$ today, we cannot have
  $\Lambda \sim H^2$ {\em always} since this would amount to a
  substantial renormalisation of $M_\mathrm{P}$ in eq.(\ref{fried})
  (taking $\kappa=0$ as suggested by inflation); this is inconsistent
  with primordial nucleosynthesis which requires the ``cosmic''
  Newton's Constant to be within a few \% of its laboratory value
  \cite{Cyburt:2004yc}. Thus arguably the most natural solution to the
  `coincidence problem' is ruled out empirically.} It has been
suggested that the `dark energy' may not be a cosmological {\em
  constant} but rather the slowly evolving potential energy $V (\phi)$
of a hypothetical scalar field $\phi$ named `quintessence' which can
{\em track} the matter energy density (see
\cite{Copeland:2006wr}). This however is equally fine-tuned since we
need $V^{1/4} \sim 10^{-12}$~GeV but $\sqrt{\mathrm{d}^2
  V/\mathrm{d}\phi^2} \sim H_0 \sim 10^{-42}$~GeV (in order that the
evolution of $\phi$ be sufficiently slowed by the Hubble expansion),
and moreover does not address the fundamental
issue,\footnote{Admittedly this criticism applies also to attempts to
  do away with dark energy by interpreting the data in terms of
  modified cosmological models, but less so since these do not invoke
  gravitating vacuum energy at all.}  namely why are all the other
possible contributions to the vacuum energy absent?  Given the no-go
theorem against any dynamical cancellation mechanism in
eq.(\ref{tuning}) in the framework of general relativity
\cite{Weinberg:1988cp}, it might appear that solving the problem will
necessarily require modification of our understanding of
gravity. Interesting suggestions have been made in this context
e.g. the DGP model in which gravity alone propagates in a new
dimension that opens up on distance scales of ${\cal O}(H_0^{-1})$
(see \cite{Lue:2005ya}). However since this is an unnaturally large
scale for a fundamental theory, this model is clearly just as fine
tuned as the cosmological constant and moreover suffers from intrinsic
theoretical difficulties such as violation of unitarity due to
`ghosts' (see \cite{Koyama:2007za}).

The situation is so desperate that `anthropic' arguments have been
advanced to explain why the cosmological constant is of just the right
order of magnitude to allow of our existence today --- if it was much
higher then galaxies would never have have formed (see
\cite{Weinberg:2000yb}). However a recent analysis
\cite{Tegmark:2005dy} shows that the most likely value from such
Bayesian arguments is in fact 20--30 times bigger than the
observationally suggested value. Moreover, one cannot claim to have a
rigorous understanding of the prior probability distribution, lacking
a theory of the cosmological constant itself. It is commonly assumed
that the prior is flat in $\Lambda$, but if it is instead flat in say
$\log(\Lambda)$, the most likely value of $\Lambda$ would be zero. A
flat prior in $\Lambda$ is indeed expected in quintessence-like models
with a very flat potential \cite{Weinberg:2000qm}, but such models are
highly fine-tuned and have little physical basis. Whereas an uniform
distribution of $\Lambda$ does arise in the 'landscape' of the large
number ($\sim 10^{500}$) of possible vacua in string theory (see
\cite{Douglas:2006es}), there is no accepted measure for how these
vacuua may come to be populated through cosmological evolution.

Given this situation, we can ask whether the observations really
require dark energy or whether this is just an artifact of
interpreting the data using an over-idealised description of the
universe. For the FRW model we see from eq.(\ref{fried}) that deducing
$\Lambda$ to be of ${\cal O}(H_0^2)$ from observations should not be
unexpected, this being its {\em natural} scale in such a model
universe (what seems quite unnatural is the {\em implied} fundamental
energy scale of $\sqrt{H_0 M_\mathrm{P}}$ since $H_0$ is so much
smaller than $M_\mathrm{P}$). From this perspective, it is easy to see
why there have been recurring claims for $\Lambda \sim H_0^2$ --- this
is effectively forced upon us by the theoretical framework in which
the data is interpreted.

\section{The observational situation}

\subsection{The age of the universe and the Hubble constant}

A recurring argument for a cosmological constant has come from
consideration of the age and present expansion rate of a FRW
cosmology. For a spatially flat universe, the present age is $t_0 =
2/3H_0$ for $\Omega_\mathrm{m} = 1$ and it has often been suggested
that the observed age is too long to be consistent with the observed
Hubble parameter for such an universe which has always been
decelerating. By contrast, an universe with $\Omega_\mathrm{m} = 0.3$,
$\Omega_\Lambda = 0.7$ would be older with $t_0 \simeq 1/H_0$ (see
\cite{Carroll:1991mt}), as is also the case for a `Milne universe'
which expands at a constant rate.

Advances in astronomical techniques now enable direct radioactive
dating using stellar spectra, e.g. the detection of singly ionized
$^{238}$U in the extremely metal-poor star CS31802-001 in the Galactic
halo implies an age of $12.5 \pm 3$~Gyr \cite{Cayrel:2001qi}. This is
consistent with the (95\% c.l.) range of $10.4 - 16$~Gyr for the age
of (the oldest stars in) globular clusters which is indirectly
inferred from the observed Hertzprung-Russell diagram using stellar
evolution modelling, after adding on $\sim 1$~Gyr, the estimated epoch
of star formation \cite{Krauss:2003em}.

With regard to the present expansion rate, the {\sl Hubble Key
  Project} \cite{Freedman:2000cf} has provided direct measurements of
the distances to 18 nearby spiral galaxies (using Cepheid variables)
and these have been used to calibrate five secondary methods (such as
Type~Ia supernovae) which probe to deeper distances and yield:
\begin{equation}
 H_0 = 72 \pm 3 \pm 7~\mathrm{km}~\mathrm{s}^{-1}~\mathrm{Mpc}^{-1}.
\label{hkp}
\end{equation}
This has come to be the generally accepted value e.g. by the Particle
Data Group \cite{Yao:2006px}. It has been argued however that the {\sl
  HKP} data need to be corrected for local peculiar motions using a
more sophisticated flow model than was actually used, and also for
metallicity (heavy element abundance) effects on the Cepheid
calibration --- this would lower $H_0$ to $63 \pm
6~\mathrm{km}~\mathrm{s}^{-1}~\mathrm{Mpc}^{-1}$
\cite{RowanRobinson:2000ys,RowanRobinson:2002iq}. The sensitivity of
the Cepheid calibration to metallicity is also emphasised in a
reanalysis of the {\sl HKP} data which finds larger distances to the 6
SNe~Ia-calibrating galaxies used by \cite{Freedman:2000cf}, thus
obtaining $H_0 = 62.3 \pm 1.3 \pm
5~\mathrm{km}~\mathrm{s}^{-1}~\mathrm{Mpc}^{-1}$
\cite{Sandage:2006cv}.

\begin{figure}[tbh]
\centering 
\includegraphics[width=0.48\textwidth]{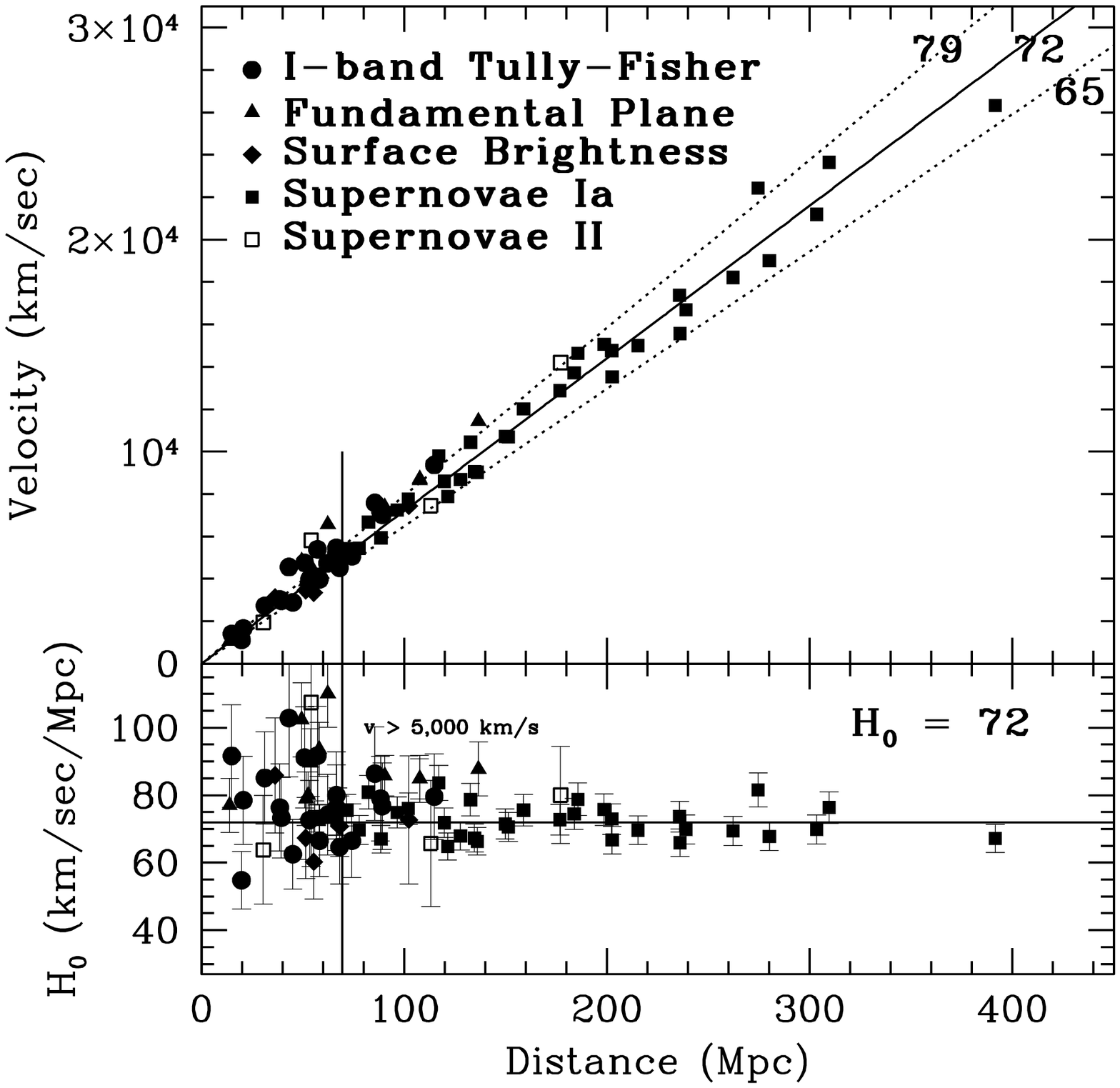}\quad
\includegraphics[width=0.48\textwidth]{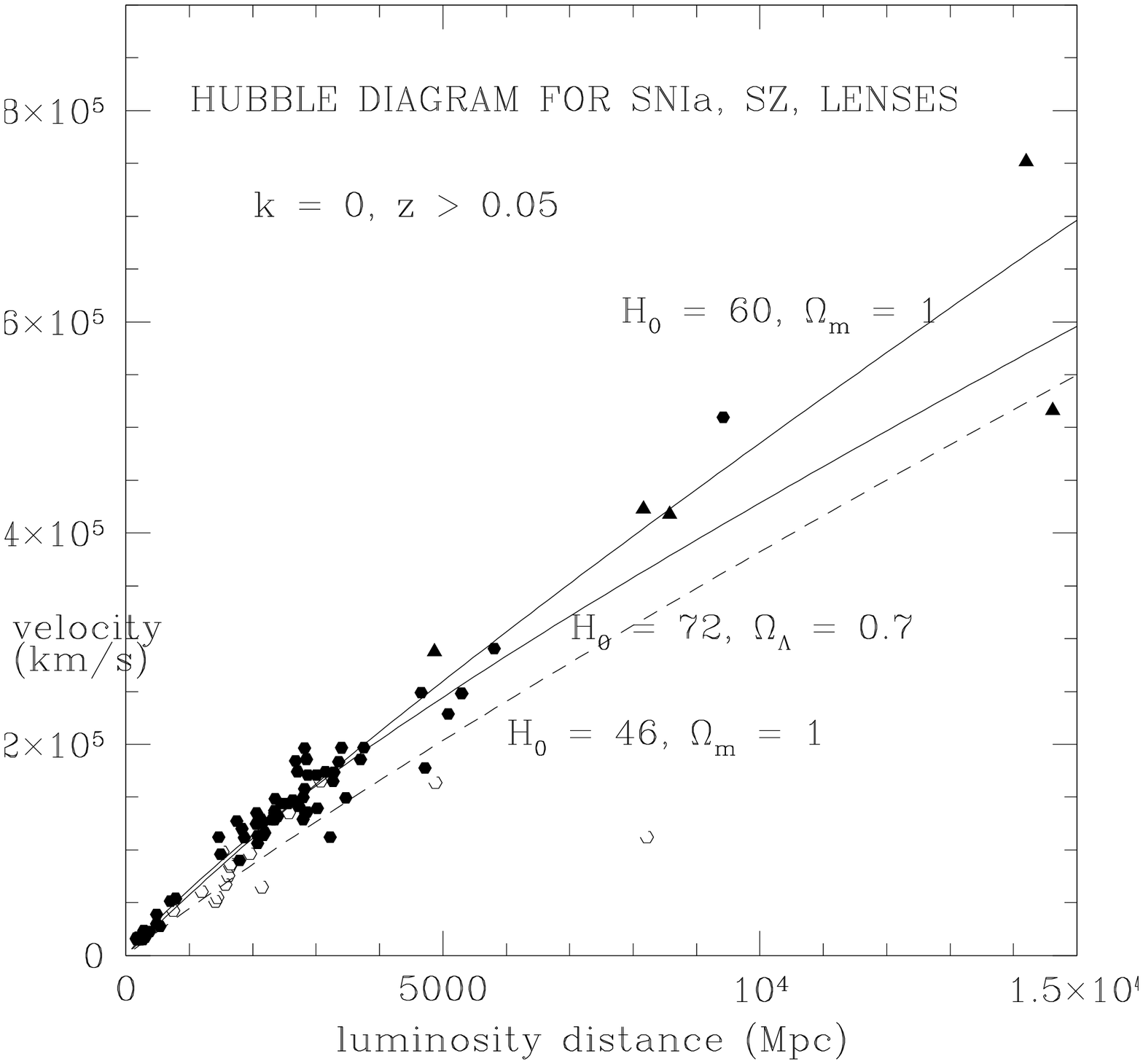}
\caption{Hubble diagram for Cepheid-calibrated secondary distance
  indicators from the {\sl Hubble Key Project} \cite{Freedman:2000cf},
  along with deeper measurements using SNe~Ia (filled circles),
  gravitational lenses (triangles) and the Sunyaev-Zeldovich effect
  (circles), with model predictions (from \cite{Blanchard:2003du}).}
\label{h0}
\end{figure}

Even smaller values of $H_0$ have been deduced using physical methods
such as measurements of time delays in gravitationally lensed systems,
which bypasses the traditional `distance ladder' calibration and
probes to far deeper distances. Using the well-measured time delays of
ten multiply-imaged quasars and taking the lenses to be isothermal
dark matter halos yields $H_0 = 48 \pm
3~\mathrm{km}~\mathrm{s}^{-1}~\mathrm{Mpc}^{-1}$
\cite{Kochanek:2003pi}; however the observations are consistent with
the {\sl HKP} value (\ref{hkp}) if there are varying amounts of dark
matter in the lensing galaxies, consistent with the mass profiles
obtained for them from stellar population evolution models
\cite{Saha:2006sw}. Measurements of the Sunyaev-Zeldovich effect in 41
X-ray emitting galaxy clusters also indicated a low value of $H_0 \sim
61 \pm 3 \pm 18~\mathrm{km}~\mathrm{s}^{-1}~\mathrm{Mpc}^{-1}$ for a
$\Omega_\mathrm{m} = 0.3$, $\Omega_\Lambda = 0.7$ universe, dropping
further to $H_0 \sim 54~\mathrm{km}~\mathrm{s}^{-1}~\mathrm{Mpc}^{-1}$
for $\Omega_\mathrm{m} = 1$ (see \cite{Reese:2003ya}). However, a
recent analysis of 38 clusters finds higher values of, respectively,
$76.9^{+3.9}_{-3.4}~\mathrm{km}~\mathrm{s}^{-1}~\mathrm{Mpc}^{-1}$ and
$67^{+4.5}_{-3.6}~\mathrm{km}~\mathrm{s}^{-1}~\mathrm{Mpc}^{-1}$ for
the two cases, with an estimated systematic uncertainty of about $\pm
10~\mathrm{km}~\mathrm{s}^{-1}~\mathrm{Mpc}^{-1}$
\cite{Bonamente:2005ct}. As seen in Figure~\ref{h0}, the systematic
uncertainties presently preclude a definitive conclusion using either
method.

Even so this raises the question whether there may be spatial
variations in the measured Hubble rate, e.g. because we are inside an
underdense region (`Hubble bubble') that is expanding faster than the
average \cite{Zehavi:1998gz}. Such voids are seen in all surveys
of large-scale structure (see e.g. \cite{Geller:1989da}) and there
are persistent indications from galaxy counts that we may be located
in such a region (e.g. \cite{Frith:2005et}). Recent work using
SNe~Ia detects a drop in the expansion rate of $\delta_H = 6.5 \pm 1.8
\%$ at a distance of 7400~km~s$^{-1}$ \cite{Jha:2006fm}),
although this is sensitive to the precise manner in which corrections
are made for reddening due to dust \cite{Conley:2007ng}. The
measured CMB dipole (due to our peculiar motion with respect to the
CMB rest frame) implies a general bound on the variance of such
deviations, if the density fluctuations are gaussian, of
${\langle\delta_H^2\rangle_R} < 10.5 h^{-1}(R/\mathrm{Mpc})^{-1}$ in a
sphere of radius $R$ \cite{Wang:1997tp}.

Moreover the {\sl HKP} data show statistically significant variations
in $H_0$ of $9~\mathrm{km}~\mathrm{s}^{-1}~\mathrm{Mpc}^{-1}$ across
the sky \cite{McClure:2007vv}. The SNe~Ia dataset also shows a
significant systematic difference in the expansion rate in the North
and South Galactic hemispheres \cite{Schwarz:2007wf}.

To conclude, current estimates of the Hubble constant are typically
quoted with 10\% uncertainty but range between 60 and
$75~\mathrm{km}~\mathrm{s}^{-1}~\mathrm{Mpc}^{-1}$ (see
\cite{Jackson:2007ug}). Thus the age argument cannot be used to
definitively exclude the Einstein--de Sitter (E-deS) model and argue
for a cosmological constant. More worryingly, there are indications
for local imhomogeneity, as well as anisotropy in the Hubble flow
which can significantly bias cosmological inferences drawn assuming an
exactly homogeneous FRW model
(e.g. \cite{Hui:2005nm,Cooray:2006ft}). It is essential that the
fundamental assumption of homogeneity underlying the standard
cosmology be rigorously tested, now that a wealth of data is available
to carry out such tests.

\subsection{The deceleration parameter}

The most exciting observational development in studies of the Hubble
expansion rate have undoubtedly been in measurements of the
deceleration parameter $q \equiv \mathrm{d} H^{-1}/\mathrm{d} t-1 =
\frac{\Omega_\mathrm{m}}{2} - \Omega_\Lambda$. This has been found to
be {\em negative} through deep studies of the Hubble diagram of SNe~Ia
pioneered by the {\sl Supernova Cosmology Project}
\cite{Perlmutter:1998np} and the {\sl High-z SN Search Team}
\cite{Riess:1998cb}. Their basic observation was that distant
supernovae at $z \sim 0.5$ are $\Delta{m} \sim 0.25$ mag
(corresponding to $10^{\Delta{m}/2.5} - 1 \simeq 25\%$) fainter than
would be expected for a decelerating universe such as the
$\Omega_\mathrm{m} = 1$ E-deS model. This has been interpreted as
implying that the expansion rate has been {\em accelerating} since
then, consequently the observed SNe~Ia are actually further away than
expected.

The measured apparent magnitude $m$ of a source of known absolute
magnitude $M$ yields the `luminosity distance':
\begin{equation}
m - M = 5\log \left(\frac{d_\mathrm{L}}{\mathrm{Mpc}}\right) + 25, \quad
d_\mathrm{L} = (1 + z) \int_0^z \frac{\mathrm{d} z'}{H (z')},
\end{equation}
which is sensitive to the expansion history, hence the cosmological
parameters. According to the second Friedmann equation (\ref{fried})
an accelerating expansion rate requires the dominant component of the
universe to have {\em negative} pressure. The more mundane alternative
possibility, namely that the SNe~Ia appear fainter because of
absorption by intervening dust, can be observationally constrained
since this would also lead to characteristic reddening, unless the
dust has unusual properties \cite{Aguirre:1998ge}. It is more
difficult to rule out that the dimming is due to evolution, i.e. that
the distant SNe~Ia (which exploded over 5 Gyr ago) are intrinsically
fainter by $\sim 25\%$ (e.g. \cite{Drell:1999dx}. Although SNe~Ia are
believed to result from the thermonuclear explosion of a white dwarf,
there is no ``standard model'' for the progenitor(s) (see
\cite{Hillebrandt:2000ga}), hence there may well be luminosity
evolution which would complicate the use of SNe~Ia as `standard
candles'.

However it is known (using nearby SNe~Ia with independently measured
distances) that their time evolution is tightly correlated with their
peak luminosities such that the intrinsically brighter ones fade
faster. This can be used to make corrections to reduce the scatter in
the Hubble diagram using various {\em empirical} methods such as a
`stretch factor' to normalise the observed apparent peak magnitudes
\cite{Perlmutter:1998np} or the `Multi-colour Light Curve Shape'
method \cite{Riess:1998cb}. Such corrections are {\em essential} to
reduce the scatter in the data sufficiently so as to allow meaningful
deductions to be made about the cosmological model. It is a matter of
concern that the corrections made by different methods do not always
correlate with each other when applied to the same objects (see
\cite{Leibundgut:2000xw}), especially since there is no physical
understanding of the observed correlations.

\begin{figure}[tbh]
\centering\leavevmode
\includegraphics[width=0.6\textwidth,angle=90]{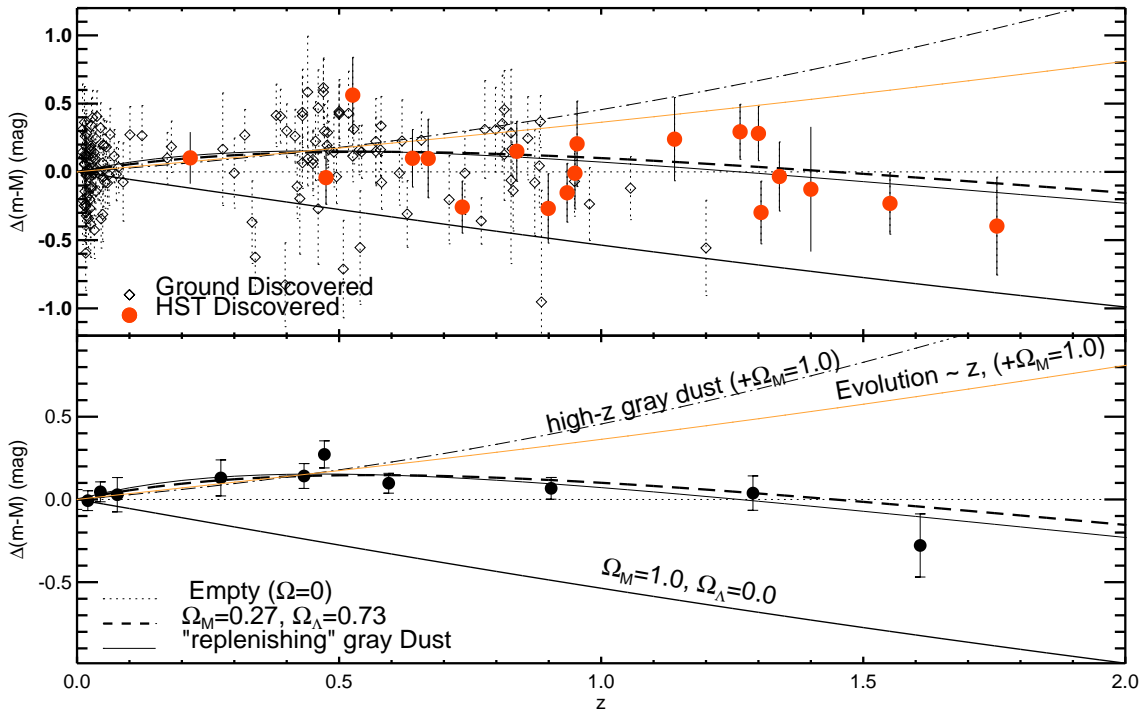}
\caption{The residual Hubble diagram of SNe~Ia relative to the
  expectation for an empty universe, compared to cosmological models;
  the bottom panel shows weighted averages in redshift bins (from
  \cite{Riess:2004nr}).}
\label{snhub}
\end{figure}

Figure~\ref{snhub} shows an example magnitude-redshift diagram of
SNe~Ia obtained by the {\sl Supernova Search Team} \cite{Riess:2004nr}
using a carefully compiled `gold set' of 142 SNe~Ia from ground-based
surveys, together with 14 SNe~Ia in the range $z \sim 1-1.75$
discovered with the {\sl Hubble Space Telescope}. The latter are {\em
  brighter} than would be expected if extinction by dust or simple
luminosity evolution ($\propto z$) is responsible for the observed
dimming of the SNe~Ia upto $z \sim 0.5$, and thus support the earlier
indication of an accelerating cosmological expansion. However
alternative explanations such as luminosity evolution proportional to
{\em lookback time}, or extinction by dust which is maintained at a
constant density are still viable. Moreover for reasons to do with how
SNe~Ia are detected, the dataset consists of approximately equal
subsamples with redshifts above and below $z \sim 0.3$. It has been
noted that this is also the redshift at which the acceleration is
inferred to begin and that if these subsets are analysed {\em
  separately}, then the 142 ground-observed SNe~Ia are consistent with
deceleration; only when the 14 high-$z$ SNe~Ia observed by the {\em
  HST} are included is there a clear indication of acceleration
\cite{Choudhury:2003tj}. Clearly further observations are necessary
particularly at the rather poorly sampled intermediate redshifts $z
\sim 0.1-0.5$ --- precisely the redshift range where the dark energy
is supposed to have come to dominate the expansion leading to
acceleration. Further observations have been made by the {\sl
  Supernova Legacy Survey} \cite{Astier:2005qq} and {\sl ESSENCE}
\cite{WoodVasey:2007jb} at the upper end of this redshift range, and
these analyses have confirmed the previous indications of accelerated
expansion.

\begin{figure}[tbh]
\centering\leavevmode
\includegraphics[width=0.58\textwidth]{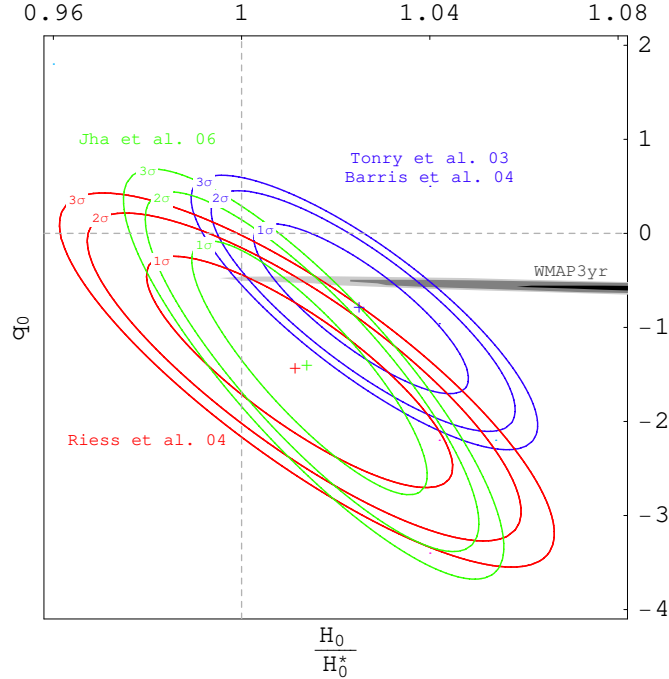}
\caption{Results from a model-independent fit to the Hubble law for
  three SNe~Ia data sets with $z < 0.2$ --- the horizontal axis shows
  the derived Hubble parameter relative to a fiducial value of $H_* =
  65~\mathrm{km}~\mathrm{s}^{-1}~\mathrm{Mpc}^{-1}$. The value of $q_0
  \simeq 0.5$ deduced from {\sl WMAP} data is shown for comparison
  (from \cite{Schwarz:2007wf}).}
\label{schwarz}
\end{figure}

A very different picture emerges however when the consistency of
different data sets is examined more closely \cite{Schwarz:2007wf}. As
mentioned earlier, the nearby ($z < 0.2$) SNe~Ia data indicates {\em
  anisotropic} expansion (at $>95\%$ c.l.), suggestive of an
unidentified systematic error. While data from the North Galactic
hemisphere are consistent with accelerated expansion, data from the
South Galactic hemisphere are not conclusive in this regard. Even when
the full-sky data is used, Figure~\ref{schwarz} shows that $q_0 = 0$
is still permitted at $2\sigma$, although $q_0 = 0.5$ (the
decelerating E-deS model) is rejected at this level. These authors
conclude: {\em ``Our model independent test fails to detect
  acceleration of the Universe at high statistical significance''}
\cite{Schwarz:2007wf}.

\section{The spatial curvature and the matter density}

Although the first indications for an accelerating universe from
SNe~Ia were rather tentative, the notion that dark energy dominates
the universe became widely accepted rather quickly
(e.g. \cite{Bahcall:1999xn}). This was because of two independent
lines of evidence which also suggested that there is a substantial
cosmological constant. The first was that contemporaneous measurements
of degree-scale angular fluctuations in the CMB by the {\sl Boomerang}
\cite{deBernardis:2000gy} and {\sl MAXIMA} \cite{Hanany:2000qf}
experiments provided a measurement of the sound horizon (a `standard
ruler') at recombination (see \cite{Hu:1996qs} and thereby indicated
that the curvature term $\kappa \simeq 0$, i.e. the universe is
spatially flat. The second was that, as had been recognised for some
time, several types of (mainly local) observations indicate that the
amount of matter which participates in gravitational clustering is
much less than the critical density, $\Omega_\mathrm{m} \simeq 0.3$
(see \cite{Peebles:2001sc}. The cosmic sum rule (\ref{sumrule}) then
requires that there be some form of `dark energy', unclustered on the
largest spatial scales probed in the measurements of
$\Omega_\mathrm{m}$, with an energy density of $1 - \Omega_\mathrm{m}
\simeq 0.7$. This was indeed consistent with the value of
$\Omega_\Lambda \sim 0.7$ suggested by the SNe~Ia data
\cite{Perlmutter:1998np,Riess:1998cb} leading to the widespread
identification of the dark energy with a cosmological constant. In
fact all data to date are consistent with $w = -1$ and this
`concordance model' is termed $\Lambda$CDM since the matter content
must mostly be cold dark matter (CDM) given the constraint from
primordial nucleosynthesis on the baryonic component
$\Omega_\mathrm{B} h^2 \simeq 0.02 \pm 0.002$ (see
\cite{Fields:2006ga}).

Subsequently a major advance has come about with precision
measurements of the CMB anisotropy by the {\sl Wilkinson Microwave
  Anisotropy Probe} \cite{Spergel:2003cb,Spergel:2006hy}, and of the
power spectrum of galaxy clustering by the {\sl 2 degree Field Galaxy
  Redshift Survey} \cite{Cole:2005sx} and the {\sl Sloan Digital Sky
  Survey} \cite{Tegmark:2003uf}. The paradigm which these measurements
test is that the early universe underwent a period of inflation which
generated a gaussian random field of of small density fluctuations
($\delta \rho/\rho \sim 10^{-5}$ with a nearly scale-invariant
`Harrison-Zeldovich' spectrum: $P (k) \propto k^n, n \simeq 1$), and
that these grew by gravitational instability in the sea of (dark)
matter to create the large-scale structure, as well as leaving a
characteristic anisotropy imprint on the `last scattering surface' of
the CMB. The latter is a snapshot of the oscillations in the coupled
baryon and photon fluids as the plasma (re)combines suddenly and the
universe becomes transparent at $z \sim 1000$ (see
\cite{Hu:1996qs}). The amplitudes and positions of the resulting
`acoustic peaks' in the angular power spectrum of the CMB are
sensitive to the cosmological parameters and it was recognised that
precision measurements of CMB anisotropy can thus be used to determine
these accurately (e.g. \cite{Bond:1993fb,Jungman:1995bz}). However in
practice there are many `degeneracies' in this exercise hence prior
assumptions have to be made concerning some of the parameters
(e.g. \cite{Bond:1997wr,Efstathiou:1998xx}). An useful analogy is to
see the generation of CMB anisotropy and the formation of LSS as a
sort of cosmic scattering experiment, in which the primordial density
perturbation is the ``beam'', the universe itself is the ``detector''
and its matter content is the ``target''. In contrast to the situation
in the laboratory, neither the properties of the beam, nor the
parameters of the target or even of the detector are known --- only
the actual ``interaction'' may be taken to be gravity. In practice
therefore assumptions have to be made about the nature of the dark
matter (e.g. `cold' non-relativistic or `hot' relativistic?) and about
the nature of the primordial perturbation (e.g. adiabatic or
isocurvature?) as well as its spectrum, together with further `priors'
(e.g. on the curvature parameter $\kappa$ or the Hubble constant $h$)
before the cosmological density parameters can be inferred from the
data. Note in particular that $\Lambda$ is dynamically quite
negligible at such large redshifts, so its only effect is to change
the distance to the last scattering surface, hence the angular scale
of the observed CMB anisotropies (in particular the position of the
first acoustic peak).

\begin{figure}[tbh]
\centering
\includegraphics[width=0.47\textwidth]{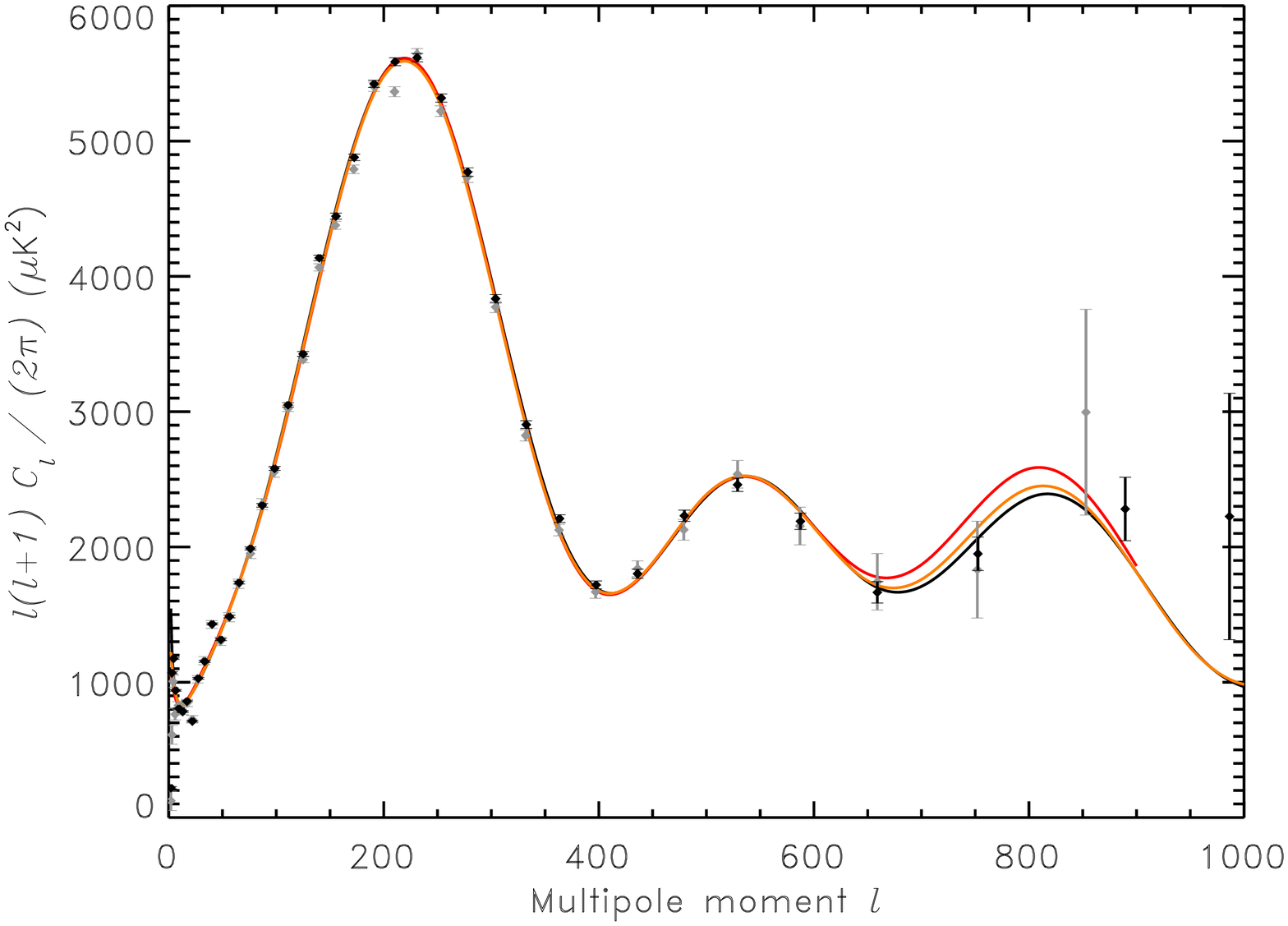}\quad
\includegraphics[width=0.44\textwidth]{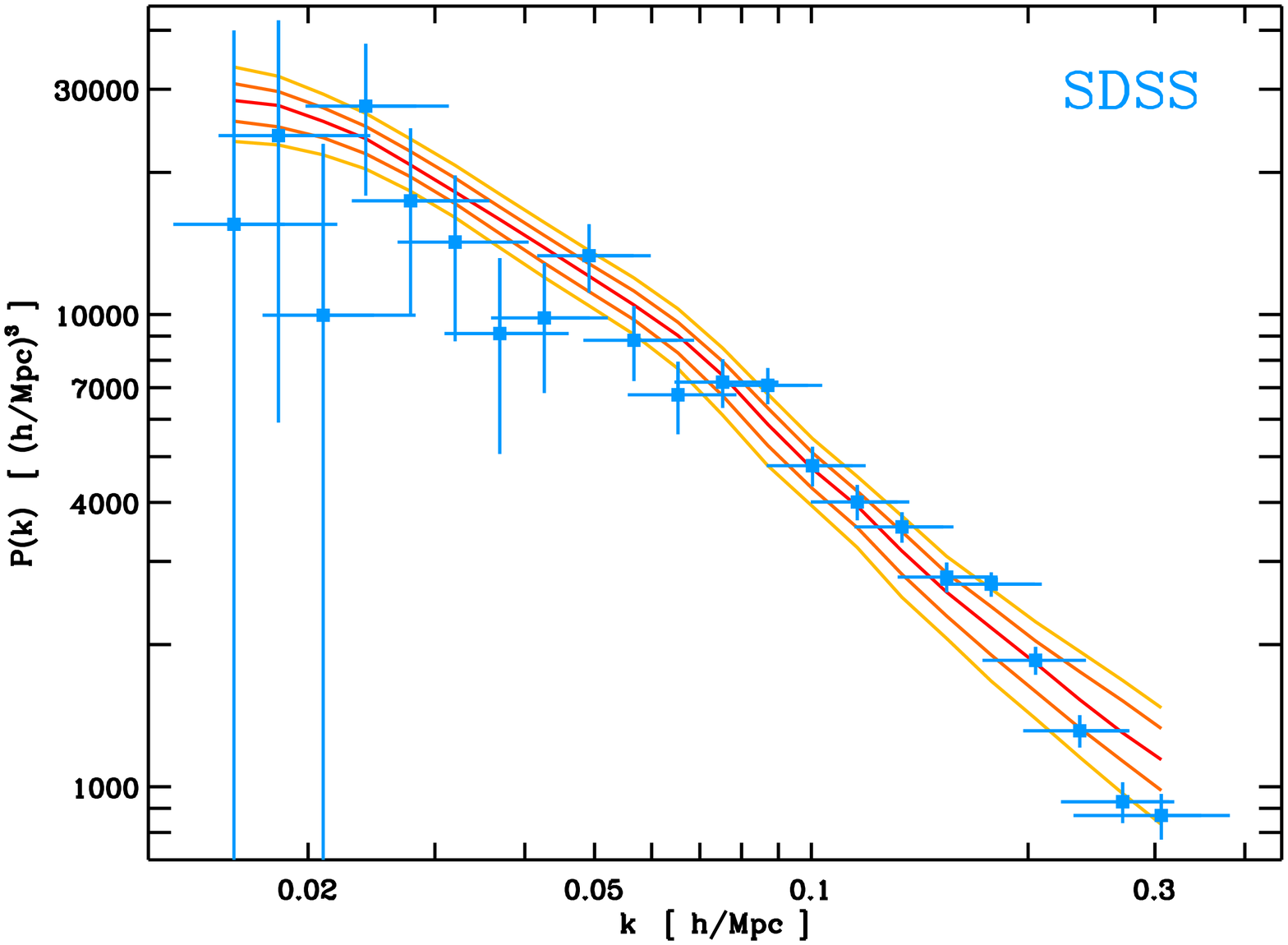}
\caption{Angular power spectrum of the CMB measured by {\sl WMAP}
  (black points -- 3 year data, gray points -- 1 year data) and the
  fit to the $\Lambda$CDM model (black line -- 3 year data, red line
  -- 1 year data); the right panel shows the predicted matter power
  spectrum compared with the {\sl SDSS} data (from
  \cite{Spergel:2006hy}).}
\label{wmap}
\end{figure}

Nevertheless as seen in Figure~\ref{wmap}, the angular spectrum of the
CMB measured by {\sl WMAP} is in impressive agreement with the
expectation for a flat $\Lambda$CDM model, assuming a power-law
spectrum for the primordial (adiabatic) perturbation
\cite{Spergel:2003cb,Spergel:2006hy}. The fitted parameters for the
3-year data are $\Omega_\mathrm{B} h^2 = 0.02229 \pm 0.00073$,
$\Omega_\mathrm{m} h^2 = 0.1277 \pm 0.008$, and $h = 0.732 \pm 0.031$
consistent with the {\sl HKP} value (\ref{hkp}). The spectral index is
obtained to be $n = 0.958 \pm 0.016$ which is as expected in simple
models of slow-roll inflation. It is particularly impressive that the
prediction for the matter power spectrum (obtained by convoluting the
primordial perturbation with the CDM `transfer function') is in
excellent agreement with the power spectrum of galaxy clustering
measured by both {\sl 2dFGRS} and {\sl SDSS}. The power spectrum from
spectral observations of the `Lyman-$\alpha$ forest' (intergalactic
gas clouds) is also concordant (e.g. \cite{Jena:2004fc}). Having
established the consistency of the $\Lambda$CDM model, this is used to
draw tight constraints e.g. on a `hot dark matter' (HDM) component
which translates into a (95\% c.l.) bound on the summed neutrino
masses of $\sum m_{\nu} < 0.66$ eV  \cite{Spergel:2006hy}.

\begin{figure}[tbh]
\centering
\includegraphics[width=0.3\textwidth, angle=-90]{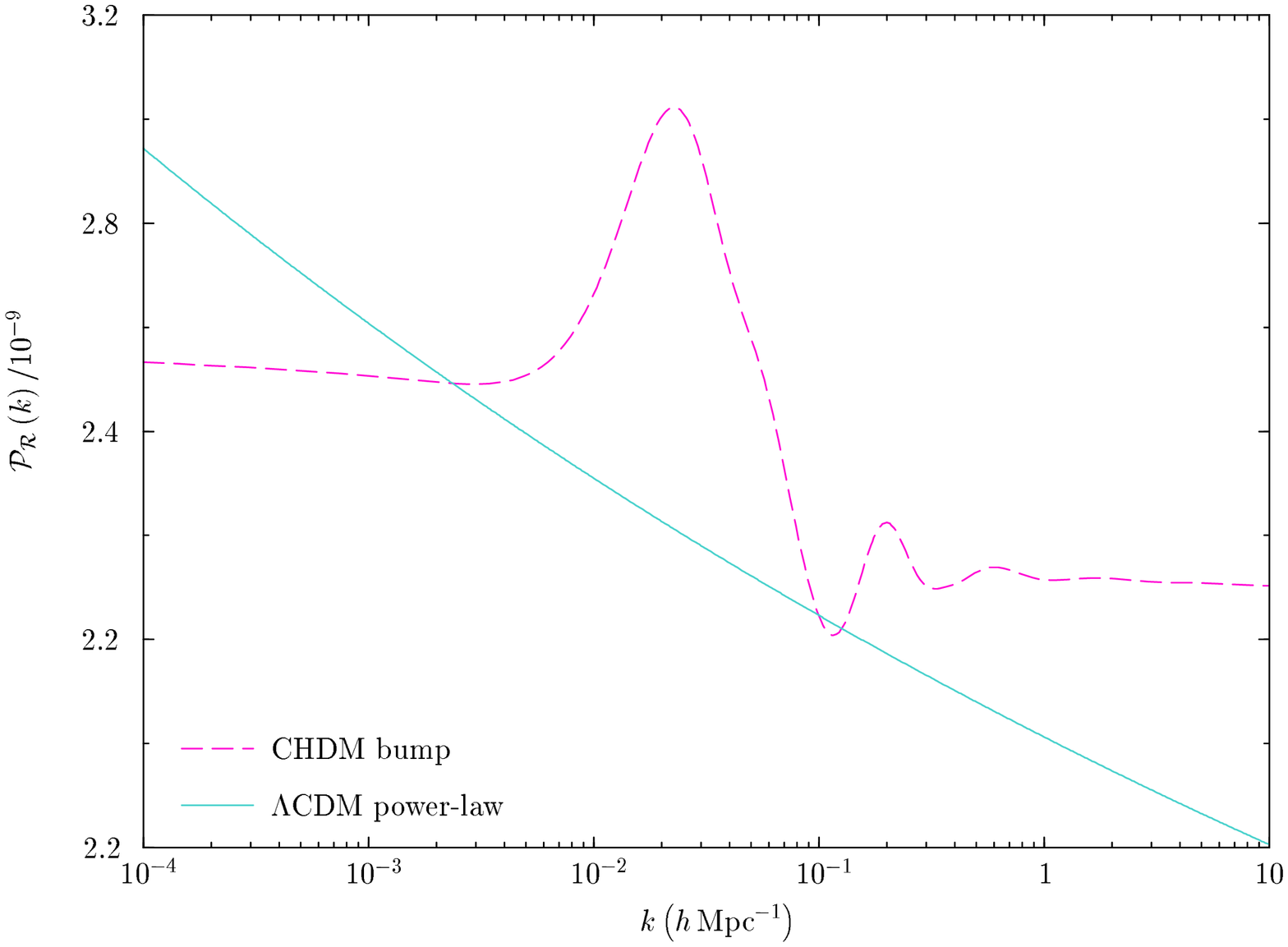}\quad
\includegraphics[width=0.3\textwidth, angle=-90]{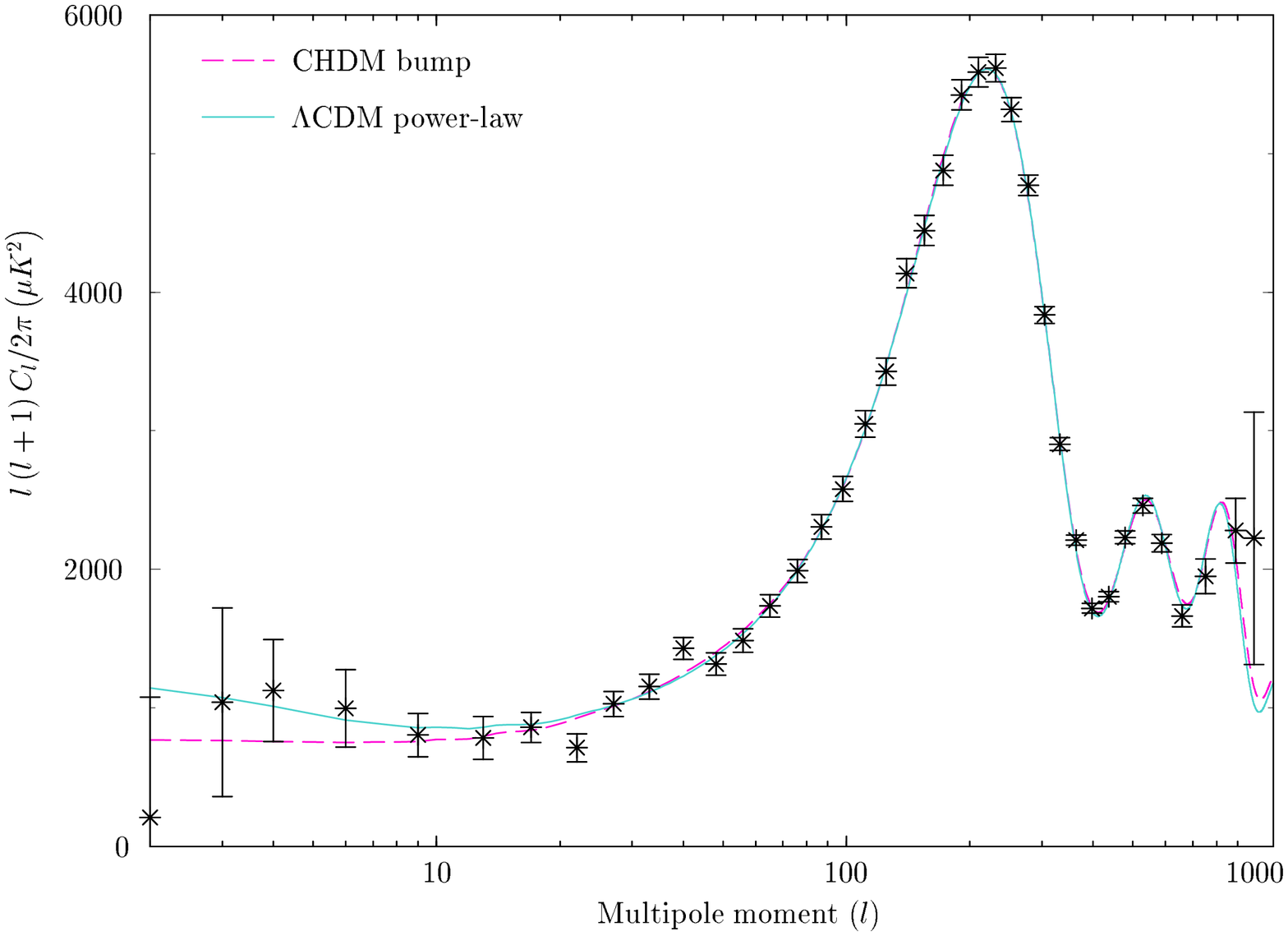}
\includegraphics[width=0.3\textwidth, angle=-90]{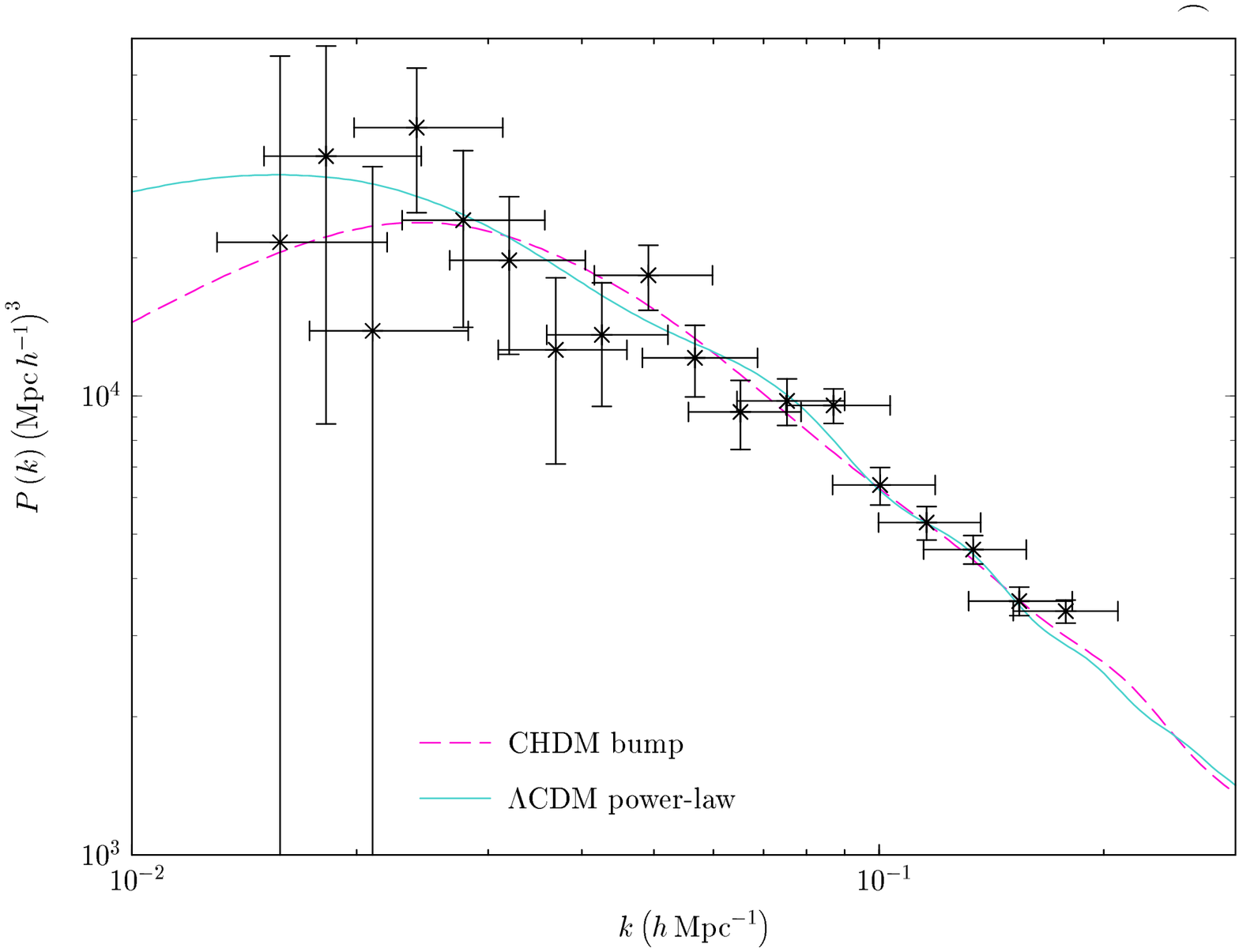}\quad
\includegraphics[width=0.3\textwidth, angle=-90]{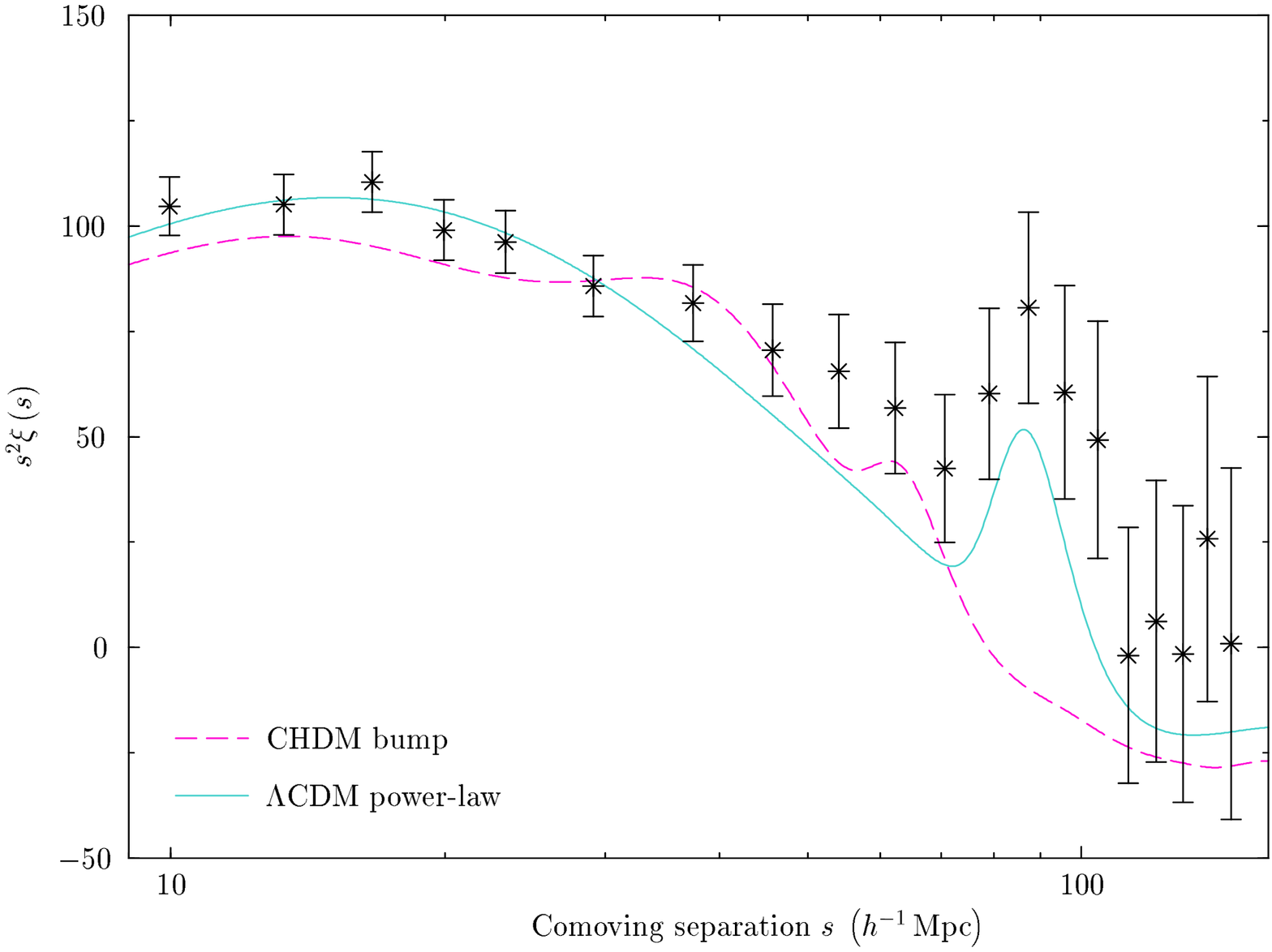}
\caption{Comparison of the power law $\Lambda$CDM model and the
  alternative CHDM model with a `bump' in the primordial spectrum (top
  left); the {\sl WMAP} data (top right) and {\sl SDSS} data (bottom
  left) are equally well-fitted, but the BAO peak in the LRG
  correlation function (bottom right) favours $\Lambda$CDM
  \cite{Hunt:2007dn}.}
\label{wmap2}
\end{figure}

It mut be pointed out however that cosmological models {\em without}
any dark energy can fit exactly the same data by making different
assumptions for the `priors'. For example, an E-deS model is still
allowed if the Hubble parameter is as low as $h \simeq 0.46$ and the
primordial spectrum is not scale-free but has a change in its slope at
a wavenumber $k \simeq 0.01$ Mpc$^{-1}$ \cite{Blanchard:2003du}. This
is a toy model but an even better fit is obtained as shown in
Figure~\ref{wmap2} with a physically motivated `bump' in the range $k
\sim (0.01 - 0.1)~h$ Mpc$^{-1}$ \cite{Hunt:2007dn}. Such a feature is
plausible in `multiple inflation' based on supergravity
\cite{Adams:1997de} wherein spontaneous symmetry breaking phase
transitions occuring during inflation create sharp changes in the mass
of the inflaton field. To satisfactorily fit the LSS power spectrum
also requires that the matter not be pure CDM but have a `hot'
component, e.g. of neutrinos with (approximately degenerate) mass
0.5~eV (i.e. $\sum m_{\nu} = 1.5$~eV) which contribute $\Omega_\nu
\simeq 0.1$. This had been noted independently for the {\sl 2dFGRS}
data \cite{Elgaroy:2003yh} and such a mass is within the sensitivity
reach of the fortcoming KATRIN $\beta$-decay experiment
\cite{Drexlin:2005zt}. It is important to note that this `cold + hot'
dark matter (CHDM) model has $\Omega_\mathrm{B}h^2 \simeq 0.02$, just
as required by primordial nucleosynthesis, but because of the low
Hubble parameter the corresponding baryon density is about 10\%. This
is quite consistent with the observed baryon fraction in X-ray
clusters, which is often used (e.g. \cite{Allen:2004cd}) to deduce
$\Omega_\mathrm{m} \sim 0.3$, hence the presence of dark energy,
adopting the {\sl HKP} prior on $h$. Moreover the value of $\sigma_8$,
the variance of mass fluctuations on cluster scales, is about $0.7$ in
this model, which is marginally consistent with measurements of cosmic
shear due to weak gravitational lensing by large-scale structure (e.g.
\cite{VanWaerbeke:2004af}).

However such a E-deS cosmology fails to match the `baryon acoustic
oscillation' (BAO) peak observed in the galaxy autocorrelation
function using luminous red galaxies at $z \sim 0.35$ in {\sl SDSS}
extending over a large survey volume. Although the measured amplitude
is below the expectation of the $\Lambda$CDM model, the position is
just as expected in this model \cite{Eisenstein:2005su}. The physical
scale of the BAO peak is independent of the cosmological model, but
its observed position (in redshift) space) is sensitive to the Hubble
parameter \cite{Blanchard:2005ev}, hence the E-deS cosmology with $h
\sim 0.5$ cannot match the data as seen in Figure~\ref{wmap2}.

\section{Conclusions}

We now have a `cosmic concordance' model with $\Omega_\mathrm{m} \sim
0.3, \Omega_\Lambda \sim 0.7$ which is supposedly consistent with all
astronomical data but has no satisfactory explanation in terms of
fundamental physics. One might hope to eventually find explanations
for the dark matter (and baryonic) content of the universe in the
context of physics beyond the Standard Model but there appears to be
little prospect of doing so for the apparently dominant component of
the universe --- the dark energy which behaves as a cosmological
constant. Cosmology has in the past been a data-starved science so it
has been appropriate to consider the simplest possible cosmological
models in the framework of general relativity. However now that we are
faced with this serious confrontation between fundamental physics and
cosmology, it would seem prudent to reconsider the underlying
assumptions, especially that of exact homogeneity.

The observed isotropy of the CMB along with a belief in the
Cosmological Principle has generally been taken to imply homogeneity
but this has come to be questioned of late. There is growing interest
in the inhomogeneous but isotropic Lemaitre-Tolman-Bondi model of a
local void (see \cite{Krasinski:1997}). It has been shown
\cite{Tomita:1999qn,Tomita:2000rf,Tomita:2000jj,Tomita:2001gh,Tomita:2002df}
that a LTB model can in principle give an explanation of the oddities
in the local Hubble flow as well as account for the SNe~Ia Hubble
diagram without invoking acceleration (see also
\cite{Celerier:1999hp,Alnes:2005rw,Enqvist:2006cg,Alnes:2006uk}). Recently
it has been claimed \cite{Biswas:2006ub} that with a large local void
of size $\sim 450h^{-1}$~Mpc the angular diameter distance of the BAO
peak can also be matched in this model. Such local inhomogeneity may
be responsible for generating the mysteriously aligned low multipoles
in the {\sl WMAP} sky \cite{Inoue:2006rd}. A void of this size does
seem extremely unlikely, but one has recently been actually detected
at $z \sim 1$ and appears to be responsible for the anomalously `cold
spot' seen by {\sl WMAP} \cite{Rudnick:2007kw}.

Landau famously said {\em ``Cosmologists are often wrong, but never in
doubt''}. The situation today is perhaps better captured by Pauli's
enigmatic remark --- the present interpretation of the data may be
{\it ``\ldots not even wrong''}. However we are certainly not without
doubt! The crisis posed by the recent astronomical observations is not
one that confronts cosmology alone; it is the spectre that haunts any
attempt to unite two of the most successful creations of 20th century
physics --- quantum field theory and general relativity. It would be fitting 
if the cosmological  constant which Einstein allegedly called his {\it 
``biggest blunder''} proves to be the catalyst for
triggering a new revolution in physics in this century.

\begin{acknowledgements}
I acknowledge a PPARC Senior Fellowship (PPA/C506205/1) and support
from the EU Marie Curie Network ``UniverseNet'' (HPRN-CT-2006-035863).
\end{acknowledgements}

\end{document}